\documentclass{sigplanconf}

\usepackage{amsmath}
\usepackage{amsfonts}
\usepackage{amssymb}
\usepackage{listings}
\usepackage{stmaryrd} 
\usepackage{amsthm}
\usepackage{thmtools}
\usepackage{units}
\usepackage{semantic}
\usepackage{tikz}
\usepackage{hyperref}
\usepackage{enumitem}

\usetikzlibrary{arrows,positioning}

\declaretheoremstyle[spaceabove=6pt, spacebelow=6pt]{thmstyle}
\declaretheorem[style=thmstyle]{definition}
\declaretheorem[style=thmstyle]{theorem}

\newcommand{\ret}{\texttt{ret }}
\newcommand{\true}{\text{true}}
\newcommand{\false}{\text{false}}
\newcommand{\sample}{\overset{\$}{\leftarrow}}

\makeatletter
\def\@copyrightspace{\relax}
\makeatother

\begin{document}

\setlength{\abovedisplayskip}{3pt}

\setlength{\pdfpageheight}{\paperheight}
\setlength{\pdfpagewidth}{\paperwidth}

\conferenceinfo{CONF 'yy}{Month d--d, 20yy, City, ST, Country} 
\copyrightyear{2014} 
\copyrightdata{978-1-nnnn-nnnn-n/yy/mm} 
\doi{nnnnnnn.nnnnnnn}





\title{The Foundational Cryptography Framework

}

\subtitle{}

\authorinfo{Adam Petcher}
          {Harvard University and MIT Lincoln Laboratory} 
           {apetcher@seas.harvard.edu}

\authorinfo{Greg Morrisett}
           {Harvard University}
           {greg@eecs.harvard.edu}


\maketitle

\begin{abstract}
We present the Foundational Cryptography Framework (FCF) for developing and checking complete proofs of security for cryptographic schemes within a proof assistant.  This is a general-purpose framework that is capable of modeling and reasoning about a wide range of cryptographic schemes, security definitions, and assumptions.  Security is proven in the computational model, and the proof provides concrete bounds as well as asymptotic conclusions.  FCF provides a language for probabilistic programs, a theory that is used to reason about programs, and a library of tactics and definitions that are useful in proofs about cryptography.  The framework is designed to leverage fully the existing theory and capabilities of the Coq proof assistant in order to reduce the effort required to develop proofs. 
\end{abstract}




\section{Introduction}

Cryptographic algorithms and protocols are becoming more numerous, specialized, and complicated.   As a result, it is likely that security vulnerabilities will slip by peer review.  To address this problem, some cryptographers~\citep{cryptoeprint:2004:331,cryptoeprint:2005:181} have proposed an increased level of rigor and formality for cryptographic proofs.  It is our hope that eventually, cryptographers will be able to describe cryptographic schemes and security proofs using a formal language, and the proofs can be checked automatically by a highly trustworthy mechanized proof checker.

To enable such mechanically-verified proofs, we have developed The Foundational Cryptography Framework (FCF).  This framework embeds into the Coq proof assistant~\cite{Coq:manual} a simple probabilistic programming language to allow the specification of cryptographic schemes, security definitions, and assumptions.  The framework also includes useful theory, tactics, and definitions that assist with the construction of proofs of security.  Once complete, the proof can be checked by the Coq proof checker.  Facts proven in FCF include the security of El Gamal encryption \cite{1057074}, and of the encryption scheme described in Section \ref{ExampleProof} of this paper.  We have also proven the security of the ``tuple-set" construction of \cite{cash13sse}, which is a significant portion of a practical searchable symmetric encryption scheme.  This is a complex and sophisticated construction, and the proof requires over 7000 lines of Coq code and includes a core argument involving more than 30 intermediate games.    

FCF is heavily influenced by CertiCrypt~\cite{Zanella:2009:POPL}, which was later followed by EasyCrypt~\cite{Zanella:2011:CRYPTO}.  CertiCrypt is a framework that is built on Coq, and allows the development of mechanized proofs of security in the computational model for arbitrary cryptographic constructions.  Unfortunately, proof development in CertiCrypt is time-consuming, and the developer must spend a disproportionate amount of time on simple, uninteresting goals.  To address these limitations, the group behind CertiCrypt developed EasyCrypt, which has a similar semantics and logic, and uses the Why3 framework and SMT solvers to improve proof automation.  EasyCrypt takes a huge step forward in terms of usability and automation, but it sacrifices some trustworthiness due to that fact that the trusted computing base is larger and the basis of the mechanization is a set of axiomatic rules.  

Following the release of EasyCrypt, a team of cryptographers and programming language experts (including one of the authors of this paper) attempted \cite{EasyCryptAPP} to prove the security of a private information retrieval system \cite{APP}.  This effort did not produce a complete proof because certain required facts could not be proven in EasyCrypt.  Specifically, it was impossible to prove particular equivalences involving loop fusion and order permutation within a loop.  In order to allow these equivalences in EasyCrypt, it would be necessary to prove them correct on paper and then modify the EasyCrypt code to include appropriate rules.  EasyCrypt has seen significant improvements since its release, and it is possible that these sorts of equivalence are now supported, but a developer may encounter some other goal that EasyCrypt does not support.  

FCF is a foundational framework like CertiCrypt, in which the rules used to prove equivalence of programs (or any fact) are mechanized proofs derived from the semantics or other core definitions.  In such a framework, the problem of the previous paragraph can be addressed by proving the appropriate theorem \emph{within} the framework, and then by using that theorem to obatain the desired equivalence.  An important difference between CertiCrypt and FCF is that CertiCrypt uses a deep embedding of a probabilistic programming language whereas FCF uses a shallow embedding (similar to \cite{cryptoeprint:2007:199}).  The shallow embedding allows us to easily extend the language, and to make better use of Coq's tactic language and existing automated tactics to reduce the effort required to develop proofs.  The result is a framework that is foundational and easily extensible, but in which proof development effort is greatly reduced.  

\section{Design Goals}

\label{properties_section}

Based on our experience working with EasyCrypt, we formulated a set of idealized design goals that a practical mechanized cryptography framework should satisfy.  We believe that FCF achieves many of these goals, though there is still some room for improvement, as discussed in Section \ref{EvaluationSection}.

\paragraph{Familiarity}
Security definitions and descriptions of cryptographic schemes should look similar to how they would appear in cryptography literature, and a cryptographer with no knowledge of programming language theory or proof assistants should be able to understand them.  Furthermore, a cryptographer should be able to inspect and understand the foundations of the framework itself.   

\paragraph{Proof Automation}
The system should use automation to reduce the effort required to develop a proof.  Ideally, this automation is extensible, so that the developer can produce tactics for solving new kinds of goals.  

\paragraph{Trustworthiness}
Proofs should be checked by a trustworthy procedure, and the core definitions (\emph{e.g.}, programming language semantics) that must be trusted in order to trust a proof should be relatively simple and easy to understand.  

\paragraph{Extensibility}
It should be possible to directly incorporate any existing theory that has been developed for the proof assistant.  For example, it should be possible to directly incorporate an existing theory of lattices in order to support cryptography that is based on lattices and their related assumptions.  

\paragraph{Concrete Security}
The security proof should provide concrete bounds on the probability that an adversary is able to defeat the scheme.  Concrete bounds provide more information than asymptotic statements, and they inform the selection of values for system parameters in order to achieve the desired level of security in practice. 

\paragraph{Abstraction}
The system should support abstraction over types, procedures, proofs, and modules containing any of these items.  Abstraction over procedures and primitive types is necessary for writing security definitions, and for reasoning about adversaries in a natural way.  The inclusion of abstraction over proofs and structures adds a powerful mechanism for developing sophisticated abstract arguments that can be reused in future proofs.

\paragraph{Code Generation}
The system should be able to generate code containing the procedures of the cryptographic scheme that was proven secure.  This code can then be used for basic testing, prototyping, or as an executable model to which future implementations will be compared during testing.    

\section{Framework Components}
\label{FrameworkSection}

In a typical cryptographic proof, we specify cryptographic schemes, security definitions, and (assumed) hard problems, and then we prove a reduction from a properly-instantiated security definition to one or more  problems that are assumed to be hard.  In other words, we assume the existence of an effective adversary against the scheme in question, and then prove that we can construct a procedure that can effectively solve a problem that is assumed to be hard.  This reduction results in a contradiction that allows us to conclude that an effective adversary against the scheme cannot exist.  

The cryptographic schemes, security definitions, and hard problems are probabilistic, and FCF provides a common probabilistic programming language (Section \ref{LangSection}) for describing all three.  Then we provide a denotational semantics (Section \ref{LangSection}) that allows reasoning about the probability distributions that correspond to programs in this language.  This semantics assigns a numeric value to an event in a probability distribution, and it also allows us to conclude that two distributions are equivalent and we can replace one with the other (which supports the game-hopping style of \citep{cryptoeprint:2004:331}).  

It can be cumbersome to work directly in the semantics, so we provide an equational theory (Section \ref{EquationalTheory}) of distributions that can be used to prove that distributions are related by equality, inequality or ``closeness."  A program logic (Section \ref{ProgramLogic}) is also provided to ease the development of proofs involving state or looping behavior.  To reduce the effort required to develop a proof, the framework provides a library of tactics (Section \ref{TacticsSection}) and a library of common program elements with associated theory (Section \ref{LibrarySection}).  The equational theory, program logic, tactics, and programming library greatly simplify proof development, yet they are all derived from the semantics of the language, and using them to complete a proof does not reduce the trustworthiness of the proof.  

By combining all of the components described above, a developer can produce a proof relating the probability that some adversary defeats the scheme to the probability that some other adversary is able to solve a problem that is assumed to be hard.  This is a result in the \textit{concrete setting}, in which probability values are given as expressions, and certain problems are assumed to be hard for particular constructed adversaries.  In such a result, it may be necessary to inspect an expression describing a probability value to ensure it is sufficiently ``small," or to inspect a procedure to ensure it is in the correct complexity class.  FCF provides additional facilities to obtain more traditional asymptotic results, in which these procedures and expressions do not require inspection.  A set of asymptotic definitions (Section \ref{AsymptoticSection}) allows conclusions like ``this probability is negligible" or ``this procedure executes a polynomial number of queries."  In order to apply an assumption about a hard problem, it may be necessary to prove that some procedure is efficient in some sense.  So FCF provides an extensible notion of efficiency (Section \ref{EfficiencySection}) and a characterization of non-uniform polynomial time Turing machines.\footnote{The current release of the FCF code for version 8.4 of Coq is included as auxiliary material.}

\subsection{Probabilistic Programs}
\label{LangSection}

We describe probabilistic programs using Gallina, the purely functional programming language of Coq, extended with a computational monad in the spirit of Ramsey and Pfeffer~\cite{Ramsey:2002:SLC:503272.503288}, that supports drawing random bit vectors from an input tape.  Listing \ref{ExampleProg} contains an example of a valid FCF program that implements a one-time pad using bit vectors.  This program accepts a bit vector argument $x$, samples a random bit vector of length $c$ (where $c$ is a constant declared outside of this function) and assigns the result to variable $p$, then returns $p \oplus x$.    

\begin{lstlisting}[float=h, captionpos=b, label=ExampleProg, caption=An Example of a Probabilistic Program]
Definition OTP (x : Bvector c) : Comp (Bvector c) 
  := p <-$ {0, 1}^c; ret (p xor x)
\end{lstlisting}

The syntax of the language is defined by an inductive type called \texttt{Comp} and is shown in Listing~\ref{CompSyntax}. At a high-level, \texttt{Comp} is an embedded domain-specific language that inherits the host language Gallina, and extends it with operations for generating and working with random bits.  

\begin{lstlisting}[float=h, captionpos=b, label=CompSyntax, caption=Probabilistic Computation Syntax]
Inductive Comp : Set -> Type :=
| Ret : forall {A : Set}{H: EqDec A}, 
    A -> Comp A
| Bind : forall {A B : Set}, 
    Comp B -> (B -> Comp A) -> Comp A
| Rnd : forall n, Comp (Bvector n)
| Repeat : forall {A : Set}, 
    Comp A -> (A -> bool) -> Comp A.
\end{lstlisting}

The most notable primitive operation is \texttt{Rnd}, which produces $n$ uniformly random bits.  The \texttt{Repeat} operation repeats a computation until some decidable predicate holds on the value returned.  This operation allows a restricted form of non-termination that is sometimes useful (\emph{e.g.}, for sampling natural numbers in a specified range).  The operations \texttt{Bind} and \texttt{Ret} are the standard monadic constructors, and allow the construction of sequences of computations, and computations from arbitrary Gallina terms and functions, respectively.  However, note that the \texttt{Ret} constructor requires a proof of decidable equality for the underlying return type, which is necessary to provide a computational semantics as seen later in this section.  In the remainder of this paper, we will use a more natural notation for these constructors: $\{0, 1\}^n$ is equivalent to \texttt{(Rnd n)}, $x \sample c; f$ is the same as \texttt{(Bind c (fun x $=>$ f)}, and \texttt{ret e} is \texttt{(Ret \_ e)}.  The framework includes an ASCII form of this notation as seen in Listing \ref{ExampleProg}.  In the case of \texttt{Ret}, the notation serves to hide the proof of decidable equality, which is irrelevant to the programmer and is usually constructed automatically by proof search.    

FCF uses a \emph{shallow embedding}, in which functions in the object language are realized using functions in the metalanguage.  In contrast, CertiCrypt uses a \emph{deep embedding}, in which the data type describing the object language includes constructs for specifying and calling functions, as well as all of the primitives such as bit-vectors and \texttt{xor}.  

We have found that there are key benefits to shallow embedding.  The primary benefit is that we immediately gain all of the capability of the metalanguage, including (in the case of Coq) dependent types, higher-order functions, modules, \emph{etc.}  Another benefit is that it is very simple to include any necessary theory in a security proof, and all of the theory that has been developed in the proof assistant can be directly utilized.  One benefit that is specific to Coq (and other proof assistants with this property) is that Gallina functions are necessarily terminating, and Coq provides some fairly complex mechanisms for proving that a function terminates.  By combining this restriction on functions with additional restrictions on \texttt{Repeat}, we can ensure that a computation (eventually) terminates, and that this computation corresponds with a distribution in which the total probability mass is $1$.  

On the other hand, the shallow embedding approach does have some drawbacks.  The main drawback is that a Gallina function is opaque; we can only reason about a Gallina function based on its input/output behavior.  The most significant effect of this limitation is that we cannot directly reason about the computational complexity of a Gallina function.  We address this issue in Section \ref{EfficiencySection}.

\begin{figure}[float = h, tb]
\begin{align*}
\llbracket \texttt{ret} \: a \rrbracket &= \textbf{1}_{\lbrace a \rbrace}\\
\llbracket x \sample c; f \: x \rrbracket &= \lambda x. \displaystyle\sum_{b \in \text{supp}(\llbracket c \rrbracket)} \left( \llbracket f \: b \rrbracket \: x \right) * \left( \llbracket c \rrbracket \: b \right) \\
\llbracket \{ 0, 1 \} ^n \rrbracket &= \lambda x. \: 2^{-n} \\
\llbracket \texttt{Repeat} \: c \: P\rrbracket &= \lambda x.
(\textbf{1}_P \: x) * (\llbracket c \rrbracket \: x) *  \left( \displaystyle\sum_{b \in P} (\llbracket c \rrbracket \: b) \right) ^ {-1}
\end{align*}
\caption{Semantics of Probabilistic Computations}
\label{DistSem}
\end{figure}

The denotational semantics of a probabilistic computation is shown in Figure \ref{DistSem}.  The denotation of a term of type \texttt{Comp A} is a function in $A \to \mathbb{Q}$ which should be interpreted as the probability mass function of a distribution on \texttt{A}.  In Figure \ref{DistSem}, $\textbf{1}_S$ is the indicator function for set $S$.  So the denotation of \texttt{ret a} is a function that returns $1$ when the argument is definitionally equal to $a$, and $0$ otherwise.  We can view the denotation of $x \sample c; f \: c$ as a marginal probability of the joint distribution formed by $c$ and $f$.  We know the probability of all events in $c$, but we only know the probability of events in $f$ conditioned on events in $c$, so we can compute the probability of any event in this marginal distribution using the law of total probability.  The fact that random bits are uniform and independent is encoded in the denotation of $\{ 0, 1 \}^n$, which is a function that ignores the argument and returns the probability that any $n$-bit value is equal to a randomly chosen $n$-bit value.  The probability that $\texttt{Repeat} \: c \: P$ produces $x$ is the conditional probability of $x$ given $P$ in $c$---which is equivalent to the function shown in Figure \ref{DistSem}.

It is important to note that this language is purely functional, but the monadic style gives programs an imperative appearance.  This appearance supports the \textit{Familiarity} design goal since cryptographic definitions and games are typically written in an imperative style.  

It is sometimes necessary to include some state in a cryptographic definition or proof. 
This can be easily accomplished by layering a state monad on top of \texttt{Comp}.  However, this simple approach does not allow the development of definitions in which an adversary has access to an oracle that must maintain some hidden state across multiple interactions with the adversary.  The definition could not simply pass the state to the adversary, because then the adversary could inspect or modify it.  So FCF provides an extension to \texttt{Comp} for probabilistic procedures with access to a stateful oracle.  The syntax of this extended language (Listing~\ref{OracleCompSyntax}) is defined in another inductive type called \texttt{OracleComp}, where \texttt{OracleComp A B C} is a procedure that returns a value of type \texttt{C}, and has access to an oracle that takes a value of type \texttt{A} and returns a value of type \texttt{B}.  

\begin{lstlisting}[float=h, captionpos=b, label=OracleCompSyntax, caption= Computation with Oracle Access Syntax]
Inductive OracleComp : 
    Set -> Set -> Set -> Type :=
| OC_Query : forall (A B : Set), 
    A -> OracleComp A B B
| OC_Run : forall (A B C A' B' S : Set), 
    EqDec S -> EqDec B -> EqDec A ->
    OracleComp A B C -> S ->
    (S -> A -> OracleComp A' B' (B * S)) ->
  OracleComp A' B' (C * S)
| OC_Ret : forall A B C, 
    Comp C -> OracleComp A B C
| OC_Bind : forall A B C C', 
    OracleComp A B C -> 
    (C -> OracleComp A B C') -> 
    OracleComp A B C'.
\end{lstlisting}

The \texttt{OC\_Query} constructor is used to query the oracle, and \texttt{OC\_Run} is used to run some program under a different oracle that is allowed to access the current oracle.  The \texttt{OC\_Bind} and \texttt{OC\_Ret} constructors are used for sequencing and for promoting terms into the language, as usual.  In the rest of this paper, we overload the sequencing and \texttt{ret} notation in order to use them for \texttt{OracleComp} as well as \texttt{Comp}.  We use \texttt{query} and \texttt{run}, omitting the additional types and decidable equality proofs, as notation for the corresponding constructors of \texttt{OracleComp}.     

\begin{figure}[float = h, tb]
\begin{align*}
\llbracket \texttt{query} \: a \rrbracket &= \lambda o \: s. ( o \: s \: a ) \\
\llbracket \texttt{run} \: c' \: o' \: s' \rrbracket &= \lambda o \: s. \llbracket c' (\lambda x \: y. \llbracket (o' (fst \: x) \: y) \: o \: (snd \: x) \rrbracket ) \: (s', s) \rrbracket \\
\llbracket \texttt{ret} \: c \rrbracket &= \lambda o \: s. x \sample c; \texttt{ret} \: (x, s) \\
\llbracket x \sample c; f \: x \rrbracket &= \lambda o \: s. [x, s'] \sample \llbracket c \: o \: s \rrbracket; \llbracket (f \: x) \: o \: s' \rrbracket
\end{align*}
\caption{Semantics of Computations with Oracle Access}
\label{OCDistSem}
\end{figure}

The denotation of an \texttt{OracleComp} is a function from an oracle and an oracle state to a \texttt{Comp} that returns a pair containing the value provided by the \texttt{OracleComp} and the final state of the oracle.  The type of an oracle that takes an \texttt{A} and returns a \texttt{B} is \texttt{(S -> A -> Comp(B * S))} for some type \texttt{S} which holds the state of the oracle.  The denotational semantics is shown in Figure \ref{OCDistSem}.  

\subsection{(In)Equational Theory of Distributions}
\label{EquationalTheory}

A common goal in a security proof is to compare two distributions with respect to some particular value (or pair of values) in the distributions.  To assist with such goals, we have provided an (in)equational theory for distributions.  This theory contains facts that can be used to show that two probability values are equal, that one is less than another, or that the distance between them is bounded by some value.  For simplicity of notation, equality is overloaded in the statements below in order to apply to both numeric values and distributions.  When we say that two distributions (represented by probability mass functions) are equal, as in $D_1 = D_2$, we mean that the functions are extensionally equal, that is $\forall x, (D_1 \: x) = (D_2 \: x)$.

\begin{theorem}[Monad Laws]
\label{MonadLaws}
\begin{align*}
\llbracket a \sample \texttt{ret } b; f a \rrbracket &= \llbracket (f \: b) \rrbracket \\
\llbracket a \sample c; \texttt{ret } a \rrbracket &= \llbracket c \rrbracket \\
\llbracket a \sample (b \sample c_1; c_2 \: b); c_3 \: a \rrbracket &= 
\llbracket b \sample c_1; a \sample c_2 \: b; c_3 \: a \rrbracket 
\end{align*}
\end{theorem}

\begin{theorem}[Commutativity]
\label{Commutativity}
$$
\begin{array}{l}
\llbracket a \sample c_1; b \sample c_2; c_3 \: a \: b \rrbracket = \llbracket b \sample c_2; a \sample c_1; c_3 \: a \: b \rrbracket
\end{array}
$$
\end{theorem}

\begin{theorem}[Distribution Irrelevance]
\label{DistributionIrrelevance}
For any well-formed computation c, 
\text{\\}
$$
\begin{array}{l}
\left( \forall x \in supp( \llbracket c \rrbracket ), \llbracket f \: x \rrbracket y = v \right) \Rightarrow 
\llbracket a \sample c; f \: a \rrbracket y = v
\end{array}
$$
\end{theorem}

\begin{theorem}[Distribution Isomorphism]
\label{DistributionIsomorphism}
For any f which is a bijection from $supp(\llbracket c_2 \rrbracket)$ to $supp( \llbracket c_1 \rrbracket)$,
\begin{align*}
& \quad \forall x \in supp(\llbracket c_2 \rrbracket ), \llbracket c_1 \rrbracket (f \: x) = \llbracket c_2 \rrbracket x \\
& \wedge
\forall x \in supp(\llbracket c_2 \rrbracket ), \llbracket f_1 \: (f \: x) \rrbracket \: v_1 = \llbracket f_2 \: x \rrbracket v_2 \\
& \Rightarrow
\llbracket a \sample c_1; f_1 \: a \rrbracket \: v_1 = \llbracket a \sample c_2; f_2 \: a \rrbracket \: v_2
\end{align*}
\end{theorem}

\begin{theorem}[Identical Until Bad]
\label{IdenticalUntilBad}
\begin{align*}
& \llbracket a \sample c_1; \ret (B \: a)\rrbracket = \llbracket a \sample c_2; \ret (B \: a)\rrbracket  \:\:\:\: \wedge  \\
& \llbracket a \sample c_1; \ret (P \: a, B \:a) \rrbracket (x, \false) = \\
& \llbracket a \sample c_2; \ret (P \: a, B \: a) \rrbracket (x, \false) \Rightarrow \\
& \:\:\:\: | \: \llbracket a \sample c_1; \ret (P \: a) \rrbracket \: x - \llbracket a \sample c_2; \ret (P \: a) \rrbracket \: x \: | \le \\
& \:\:\:\:  \llbracket a \sample c_1; \ret (B \: a) \rrbracket \: \true
\end{align*}
\end{theorem}

The meaning and utility of many of the above theorems is direct (such as the standard monad properties in Theorem \ref{MonadLaws}), but others require some explanation.  Theorem \ref{DistributionIrrelevance} considers a situation in which the probability of some event $y$ in $\llbracket f \: x \rrbracket$ is the same for all $x$ produced by computation $c$.  Then the distribution $\llbracket c \rrbracket$ is irrelevant, and it can be ignored.  This theorem only applies to \textit{well-formed} computations: A well-formed computation is one that terminates with probability $1$, and therefore corresponds to a valid probability distribution.        

Theorem \ref{DistributionIsomorphism} is a powerful theorem that corresponds to the common informal argument that two random variables ``have the same distribution."  More formally, assume distributions $\llbracket c_1 \rrbracket$ and $\llbracket c_2 \rrbracket$ assign equal probability to any pair of events $(f \: x)$ and $x$ for some bijection $f$.  Then a pair of sequences beginning with $c_1$ and $c_2$ are denotationally equivalent as long as the second computations in the sequences are equivalent when conditioned on $(f \: x)$ and $x$.  A special case of this theorem is when $f$ is the identity function, which allows us to simply ``skip" over two semantically equivalent computations at the beginning of a sequence. 

Theorem \ref{IdenticalUntilBad}, also known as the ``Fundamental Lemma" from \citep{cryptoeprint:2004:331}, is typically used to bound the distance between two games by the probability of some unlikely event.  Computations $c_1$ and $c_2$ produce both a value of interest and an indication of whether some ``bad" event happened.  We use (decidable) predicate $B$ to extract whether the bad event occurred, and projection $P$ to extract the value of interest.  If the probability of the ``bad" event occurring in $c_1$ and $c_2$ is the same, and if the distribution of the value of interest is the same in $c_1$ and $c_2$ when the bad event does not happen, then the distance between the probability of the value of interest in $c_1$ and and $c_2$ is at most the probability of the ``bad" event occurring.

\subsection{Program Logic}
\label{ProgramLogic}

The final goal of a cryptographic proof is always some relation on probability distributions, and in some cases it is possible to complete the proof entirely within the equational theory described in \ref{EquationalTheory}.  However, when the proof requires reasoning about loops or state, a more expressive theory may be needed in order to discharge some intermediate goals.  For this reason, FCF includes a program logic that can be used to reason about changes to program state as the program executes.  Importantly, the program logic is related to the theory of probability distributions through completeness and soundness theorems which allow the developer to derive facts about distributions from program logic facts, and vice-versa.  

The core logic is a Probabilistic Relational Postcondition Logic (PRPL), that behaves like a Hoare logic, except there are no preconditions.  The definition of a PRPL specification is given in Definition \ref{PRPLSpec}.  In less formal terms, we say that computations $p$ and $q$ are related by the predicate $\Phi$ if both $p$ and $q$ are marginals of the same joint probability distribution, and $\Phi$ holds on all values in the support of that joint distribution.  

\begin{definition}[PRPL Specification] 
\label{PRPLSpec}
Given $p$ : \texttt{Comp A} and $q$ : \texttt{Comp B}, $p \sim q \{\Phi\}$ iff,
\begin{align*}
& \exists \: (d : \texttt{Comp (A * B)}), \forall (x, y) \in supp(\llbracket d \rrbracket ), \Phi \: x \: y \: \wedge \\
& \llbracket p \rrbracket = \llbracket x \sample d; \ret (fst \: x) \rrbracket \wedge 
 \llbracket q \rrbracket = \llbracket x \sample d; \ret (snd \: x) \rrbracket 
\end{align*}
\end{definition}

Using the PRPL, we can construct a Probabilistic Relational Hoare Logic (PRHL) which includes a notion of precondition for functions that return computations as shown in Definition \ref{PRHLSpec}.  
The resulting program logic is very similar to the Probabilistic Relational Hoare Logic of EasyCrypt~\cite{Zanella:2011:CRYPTO}, and it has many of the same properties.     

\begin{definition}[PRHL Specification] 
\label{PRHLSpec}
Given $p$ : \texttt{A -> Comp B} and $q$ : \texttt{C -> Comp D}, $\{\Psi\} p \sim q \{\Phi\}$ iff, $\forall a \: b, \Psi \: a \: b \Rightarrow (p \: a) \sim (q \: b) \{\Phi\}$.
\end{definition}

Several theorems are provided along with the program logic definitions to simplify reasoning about programs.  In order to use the program logic, one only needs to apply the appropriate theorem, so it is not necessary to produce the joint distribution described in the definition of a PRPL specification unless a suitable theorem is not provided.  Theorems are provided for reasoning about the basic programming language constructs, interactions between programs and oracles, specifications describing equivalence, and the relationship between the program logic and the theory of probability distributions.  Some of the more interesting program logic theorems are described below.  

\begin{theorem}[Soundness/Completeness w.r.t. Equality]
\label{EqualitySoundComplete}
$$
p \sim q \{\lambda \: a \: b. a = x \Leftrightarrow b = y \}  \Leftrightarrow 
\llbracket p \rrbracket \: x = \llbracket q \rrbracket  \: y
$$
\end{theorem}

\begin{theorem}[Soundness/Completeness w.r.t. Inequality]
\label{InequalitySoundComplete}
$$
p \sim q \{\lambda \: a \: b. a = x \Rightarrow b = y \}  \Leftrightarrow 
\llbracket p \rrbracket \: x \le \llbracket q \rrbracket  \: y
$$
\end{theorem}

\begin{theorem}[Sequence Rule]
\label{PRHLSeq}
\begin{align*}
p \sim q \{\Phi' \} \Rightarrow \{\Phi' \} r \sim s \{\Phi \} \Rightarrow \\
(x \sample p; r \: x) \sim (x \sample q; s \: x) \{\Phi \}
\end{align*}
\end{theorem}

\begin{theorem}[Oracle Equivalence]
\label{OracleEquiv}
Given an \textit{OracleComp} $c$, and a pair of oracles, $o$ and $p$ with initial states $s$ and $t$,
\begin{align*}
& \Phi = \lambda \: x \: y. (fst \: x) = (fst \: y) \wedge P \: (snd \: x) (snd \: y)  \Rightarrow \\
& \left( \forall a \: s' \: t', P \: s' \: t' \Rightarrow (o \: s' \: a) \sim (p \: t' \: a) \{\Phi \} \right)   \Rightarrow \\
& P \: s \: t \Rightarrow (\llbracket c \rrbracket \: o \: s) \sim (\llbracket c \rrbracket \: p \: t) \{ \Phi \}
\end{align*}
\end{theorem}

Theorems \ref{EqualitySoundComplete} and \ref{InequalitySoundComplete} relate judgments in the program logic to relations on probability distributions.  The forward direction (soundness) is typically used in a proof to transform the goal into the program logic in order to accurately reason about loops and/or state.  Once the goal is in the program logic, the backward direction (completeness) can be used to return to a goal about distributions, or to apply an existing theorem that describes relations on probability distributions.  Theorem \ref{PRHLSeq} is the relational form of the standard Hoare logic sequence rule, and it supports the decomposition of program logic judgments.  

Theorem \ref{OracleEquiv} allows the developer to replace some oracle with an observationally equivalent oracle.  This theorem takes a relational invariant $P$ on the states of the oracles, and requires the developer to prove that if $P$ holds on the states of the oracles and they are given identical input, then $P$ holds on the resulting states and the outputs are identical.  As long as $P$ holds on the initial state of the oracle, this theorem concludes that the values returned by the program interacting with the oracles are equal, and that $P$ holds on the final state.  There is also a more general form of this theorem (omitted for brevity) in which the state of the oracle is allowed to go bad, and the interaction only produces equivalent results if the state does not go bad.  This more general theorem can be combined with Theorem \ref{IdenticalUntilBad} to get ``identical until bad" results for program/oracle interactions.  

\subsection{Tactics}
\label{TacticsSection}
The framework includes several tactics that can be used to transform goals using the facts in Sections \ref{EquationalTheory} and \ref{ProgramLogic}.  For example, the \texttt{comp\_swap\_l} tactic applies commutativity (Theorem \ref{Commutativity}) to swap two independent statements at the beginning of the program on the left (of the equality, inequality, or program logic specification) in the current goal.  There are similar tactics for manipulating games based on Left Identity (\texttt{comp\_ret}), Associativity (\texttt{comp\_inline}), Distribution Irrelevance (\texttt{comp\_irr}), and the special case of Distribution Isomorphism in which the bijection is the identity function (\texttt{comp\_skip}).  Many of these tactics can be applied to goals related to probability distributions as well as goals in the program logic.          

A tactic called \texttt{dist\_compute} is provided to automatically discharge goals involving simple computations for which the corresponding distribution obviously has some desired property---typically that the probability of some event equals some specific value.  A common proof technique is to develop a program in which the probability value of a particular event is obvious, and then relate other programs to this one by equivalence proofs.  Then \texttt{dist\_compute} can be used to automatically compute the desired probability value for this program.  The tactic works by producing an arithmetic expression from the computation(s) and then performing case splits in appropriate ways in order to get goals that can be solved automatically by existing Coq decision procedures (such as \texttt{intuition} and \texttt{omega}).    

\subsection{Programming Library}
\label{LibrarySection}

The framework includes a library containing useful programming structures and their related theory.  For example, the library includes several sampling routines, such as drawing a natural number from a specified range; drawing an element from a finite list, set, or group; or sampling an arbitrary Bernoulli distribution.  These sampling routines are all computations based on the \texttt{Rnd} statement provided by the language, and each routine is accompanied by a theory establishing that the resulting distribution is correct and has the desired properties.   

The \texttt{CompFold} package contains \emph{higher-order} functions for folding and mapping a computation over a list.  This package uses the program logic extensively, and many of the theorems take a specification on a pair of computations as an argument, and produce a specification on the result of folding/mapping those computations over a list.  The package also contains theorems about typical list and loop manipulations such as appending, flattening, fusion/fission and order permutation.

\subsection{Asymptotic Theory}
\label{AsymptoticSection}

The bulk of the effort in a security proof will be spent obtaining some result in the concrete setting.  From there, a little more effort is required to produce a proof of some asymptotic fact that one would typically encounter in cryptography literature.  To enable such asymptotic definitions and proofs, FCF includes a library of standard asymptotic definitions such as Definitions \ref{Polynomial} and \ref{Negligible}.  The library also includes theorems that can be used to prove that functions are polynomial or negligible based on their composition(\emph{e.g.}, the sum of polynomials is polynomial, the quotient of polynomial and exponential is negligible).

\begin{definition}[At Most Polynomial]
\label{Polynomial}
A function $f : \mathbb{N} -> \mathbb{N}$ is \textit{at most polynomial} iff 
$\exists x \: c_1 \: c_2, \forall n, f(n) \le c_1 * n^x + c_2$
\end{definition}

\begin{definition}[Negligible Function]
\label{Negligible}
A function $f : \mathbb{N} -> \mathbb{Q}$ is \textit{negligible} iff 
$\forall c, \exists n, \forall x > n, f(x) < \nicefrac{1}{x^c}$
\end{definition}

\subsection{Efficient Procedures}
\label{EfficiencySection}

A typical asymptotic security property states that a family of cryptographic schemes has some desirable property for all efficient adversaries.  So in order to prove and apply these properties, we require some notion of ``efficient" (families of) procedures.  The language of computations used in FCF does not imply any particular model of computation---it is just a mechanism to specify probability distributions in a computational manner.  Any notion of ``efficiency" must first fix a model of computation, and then a complexity class on that model.  We want this notion of efficiency to be flexible and extensible, so we can support several different models of computation and complexity classes.      

To accomplish this flexibility, we parameterize asymptotic security definitions by an ``admissibility predicate" indicating the class of adversaries against which a problem is assumed to be hard, or a scheme is proven to be secure.  In this setting, the adversary is a family of procedures indexed by a natural number which indicates the value of the security parameter.  The admissibility predicate can describe the efficiency of the adversary as well as other properties such as well-formedness or the number of allowed oracle queries as a function of the security parameter.  

FCF includes a simple cost model and an associated admissibility predicate describing non-uniform worst-case polynomial time Turing machines that perform a (worst case) polynomial number of oracle queries.  This admissibility predicate is constructed using a concrete cost model that assigns numeric costs to particular Coq functions, \texttt{Comp} values, and \texttt{OracleComp} values.  In this cost model, the cost of executing a function is in $\mathbb{N}$, indicating the worst-case (over all arguments) execution time.  The cost of running a \texttt{Comp} is in $\mathbb{N}$, indicating the worst-case execution time over all outcomes.  The cost of executing an \texttt{OracleComp} is in $\mathbb{N} -> \mathbb{N}$, and is a function from the cost of executing the oracle to the cost of executing the computation, including the cost of executing all oracle queries.  

The cost model for Gallina functions is axiomatic, as there is no direct way to capture such an intensional property for these terms.  Our cost model includes axioms for primitive operations as well as a set of combinators for building more complicated functions.  For example, the model includes an axiom stating that the \texttt{xor} operation for bit vectors of length $c$ has a cost of $c$.  As other examples, the model includes axioms stating that the cost of $f$ composed with $g$ is the sum of the costs of $f$ and $g$, and the cost of $\texttt{if}\,\,e_1\,\,\texttt{then}\,\,e_2\,\,\texttt{else}\,\,e_3$ is the cost of $e_1$ plus the maximum of the costs of $e_2$ and $e_3$.  
Obviously, our cost axioms are incomplete, but in practice, the number required is relatively small since it is only necessary to reason about the functions used by a constructed adversary in a proof.  Of course, the axioms need to be carefully inspected to ensure they accurately describe the desired complexity class.\footnote{Furthermore, a proof that a Gallina term has a cost described by these axioms does not mean that the extracted OCaml code will have this complexity, but rather, there exists some (propositionally) equivalent term which has the described cost.  Since we are only trying to show the existence of an effective procedure, this is sufficient for our purposes.}  But of course, a similar kind of inspection is needed to ensure the faithfullness of a cost model for a deeply-embedded language.

\subsection{Code Extraction}
\label{extraction_section}
FCF provides a code extraction mechanism that includes a strong guarantee of equivalence between a model of a probabilistic program and the code extracted from that model.  The denotational semantics of probabilistic computations relates a computation to a probability distribution, but it does not contain sufficient information to allow us to reason about the behavior of such computations on a traditional computer.  So we developed a small-step operational semantics that describes the behavior of these computations on a machine in which the memory contains values rather than probability distributions.  The operational semantics (omitted for brevity) is an oracle machine that is given a finite list of bits representing the ``random" input, and it describes how a computation takes a single step to produce a new computation, a final value, or fails due to insufficient input bits. 

To show that this semantics is correct, we consider $[c]_n$, the multiset of results obtained by running a program $c$ under this semantics on the set of all input lists of length $n$.  We can view $[c]_n$ as a distribution, where the mass of some value $a$ in the distribution is the proportion of input strings that cause the program to terminate with value $a$.  In order to compare the operational semantics with the denotational semantics, we want to view the operational semantics as a relation between computations and distributions.  So the distribution related to computation $c$ by the operational semantics is $\lim_{n \to \infty} [ c ]_n$.  The statement of equivalence between the semantics is shown in Theorem \ref{AdequacyTheorem}. 

\begin{theorem}
\label{AdequacyTheorem}
If c is well-formed, then $$\displaystyle\lim_{n \to \infty} [ c ]_n = \llbracket c \rrbracket$$      
\end{theorem}

FCF contains a proof of Theorem \ref{AdequacyTheorem} as a validation of the operational semantics used for extraction, but this theorem also provides other benefits.  Because limits are unique, if two programs are equivalent under the operational semantics, then they are also equivalent under the denotational semantics.  This allows us to prove equivalence of two programs using the operational semantics when it is more convenient to do so.  For example, theorems related to unrolling \texttt{Repeat} statements are trivial to prove under the operational semantics.

Another benefit of the operational semantics and proof of equivalence is that this semantics can be considered to be the basic semantics for computations, and the denotational semantics no longer needs to be trusted.  Some may prefer this arrangement, since the operational semantics more closely resembles a typical model of computation, and may be easier to understand and inspect.  The operational semantics can also be used as a basis for a model of computation used to determine whether programs are efficient.  
            
Now that we have an operational semantics, we can simply use the standard Coq extraction mechanism to extract it along with the model of interest and all supporting types and functions.  Of course, the trustworthiness of the extracted code depends on the correctness of Coq's extraction mechanism.  Gallina does not allow infinite recursion, so the framework includes OCaml code that runs a computation under the operational semantics until a value is obtained.  The final step is instantiating any abstract types and functions with appropriate OCaml code.  This extraction mechanism does not produce production-quality code, but the code could be used for purposes related to prototyping and testing.

\section{Security Proof Construction}
\label{ExampleProof}

This section uses an example to describe the process of constructing a proof of security using the general process described at the beginning of Section \ref{FrameworkSection}.  We consider a simple encryption scheme constructed from a pseudorandom function (PRF), and we prove that ciphertexts produced by this scheme are indistinguishable under chosen plaintext attack (IND-CPA).  These security definitions, and the formal description of the construction, are provided in later sub-sections.  

This example proof is relatively simple, yet it contains many elements that one would find in a typical cryptographic argument, and so it allows us to exercise all of the key functionality of the framework.  A more complex mechanized proof (\emph{e.g.}, the proof of \cite{cash13sse}) may have more intermediate games and a different set of arguments to justify game transformations, but the structure is similar to the proof that follows.  

\subsection{Concrete Security Definitions}

In FCF, security definitions are used to describe properties that some construction is proven to have, as well as problems that are assumed to be hard.  In the PRF encryption proof, we use the definition of a PRF to assume that such a PRF exists, and we use that assumption to prove that the construction in question has the IND-CPA property.  A concrete security definition typically contains some game and an expression that describes the \textit{advantage} of some adversary -- \emph{i.e.}, the probability that the adversary will ``win" the game.

The game used to define the concrete security of a PRF is shown in Listing \ref{PRF_Def_Concrete}.  Less formally, we say that \textit{f} is a PRF for some adversary \textit{A}, if \textit{A} cannot effectively distinguish \textit{f} from a random function.  So this means that we expect that \texttt{PRF\_Advantage} is ``small" as long as \texttt{A} is an admissible adversary.   

\begin{lstlisting}[float=h, captionpos=b, label=PRF_Def_Concrete, caption=PRF Concrete Security Definition]
Variable Key D R : Set.
Variable RndKey : Comp Key.
Variable RndR : Comp R.
Variable A : OracleComp D R bool.
Variable f : Key -> D -> R.
    
Definition f_oracle(k : Key)
  (x : unit)(d : D) : Comp (R * unit) :=
  ret (f k d, tt).
    
Definition PRF_G_A : Comp bool := 
  k <-$ RndKey;
  [b, _] <-$2 A (f_oracle k) tt;
  ret b.
    
Definition PRF_G_B : Comp bool := 
  [b, _] <-$2 A (RndR_func) nil;
  ret b.
    
Definition PRF_Advantage := 
  | Pr[PRF_G_A] - Pr[PRF_G_B] |.  

\end{lstlisting}

The function \texttt{f\_oracle} simply puts the function \texttt{f} in the form of an oracle, though a very simple one with no state and with deterministic behavior.  The procedure \texttt{RndR\_func} is an oracle implementing a random function constructed using the provided computation \texttt{RndR}.  The expressions involving \texttt{A} use a coercion in Coq to invoke the denotational semantics for \texttt{OracleComp}, and therefore ensure that \texttt{A} can query the oracle but has no access to the state of the oracle.  

At a high level, this definition involves two games describing two different ``worlds" in which the adversary may find himself.  In one world (\texttt{PRF\_G\_A}) the adversary interacts with the PRF, and in the other (\texttt{PRF\_G\_B}) the adversary interacts with a random function.  In each game, the adversary interacts with the oracle and then outputs a bit.  The advantage of the adversary is the difference between the probability that he outputs $1$ in world \texttt{PRF\_G\_A} and the probability that he outputs $1$ in world \texttt{PRF\_G\_B}.  If \texttt{f} is a PRF, then this advantage should be small.  

The concrete security definition for IND-CPA encryption is shown in Listing \ref{IND_CPA_Def_Concrete}.  In this definition, \texttt{KeyGen} and \texttt{Encrypt} are the key generation and encryption procedures.  The adversary comprises two procedures, \texttt{A1} and \texttt{A2} with different signatures, and the adversary is allowed to share arbitrary state information between these two procedures.  This definition uses a slightly different style than the PRF definition---there is one game and the ``world" is chosen at random within that game.  Then the adversary attempts to determine which world was chosen.          

\begin{lstlisting}[float=h, captionpos=b, label=IND_CPA_Def_Concrete, caption=IND-CPA Concrete Security Definition]
Variable Plaintext Ciphertext Key State : Set.
Variable KeyGen : Comp Key.
Variable Encrypt : Key -> Ciphertext -> 
  Comp Plaintext.
Variable A1 : OracleComp Plaintext Ciphertext 
  (Plaintext * Plaintext * State).
Variable A2 : State -> Ciphertext -> 
  OracleComp Plaintext Ciphertext bool.

Definition EncryptOracle
  (k : Key)(x : unit)(p : Plaintext) :=
  c <-$ Encrypt k p;
  ret (c, tt).
 
Definition IND_CPA_SecretKey_G :=
  key <-$ KeyGen ;
  [b, _] <-$2 
  (
    [p0, p1, s_A] <--$3 A1;
    b <--$$ {0, 1};
    pb <- if b then p1 else p0;
    c <--$$ Encrypt key pb;
    b' <--$ A2 s_A c;
    $ ret eqb b b'
  )
  (EncryptOracle key) tt;
  ret b.
    
Definition IND_CPA_SecretKey_Advantage := 
  | Pr[IND_CPA_SecretKey_G] - 1 / 2 |.

\end{lstlisting}

In Listing \ref{IND_CPA_Def_Concrete}, the game produces an encryption oracle from the \texttt{Encrypt} function and a randomly-generated encryption key.  Then the remainder of the game, including the calls to \texttt{A1} and \texttt{A2}, may interact with that oracle.  The code for this definition includes some additional notation (different arrows and extra \texttt{\$} symbols) that is only used to provide hints to the Coq parser and does not change the behavior of the program.       

\subsection{Construction}
The construction, like the security definitions, can be modeled in a very natural way.  Of course, one must take care to ensure that the construction has the correct signature as specified in the desired security property.  The PRF encryption construction is shown in Listing \ref{PRF_Enc_Cons}.  

\begin{lstlisting}[float=h, captionpos=b, label=PRF_Enc_Cons, caption=Encryption using a PRF]

Variable eta : nat.
Variable f : Bvector eta -> 
  Bvector eta -> Bvector eta.

Definition PRFE_KeyGen := 
  {0, 1} ^ eta.
  
Definition PRFE_Encrypt 
  (k : Key )(p : Plaintext) :=
  r <-$ {0, 1} ^ eta;
  ret (r, p xor (f k r)).
  
Definition PRFE_Decrypt 
  (k : Key)(c : Ciphertext) :=
  (snd c) xor (f k (fst c)).

\end{lstlisting}

In the PRF Encryption construction, we assume a \texttt{nat} called \texttt{eta} ($\eta$) which will serve as the security parameter.  The encryption scheme is based on a function \texttt{f}, and the scheme will only be secure if \texttt{f} is a PRF.  The type of keys and plaintexts is bit vectors of length \texttt{eta}, and the type of ciphertexts is pairs of these bit vectors.  The decryption function is included for completeness, but it is not needed for this security proof.    

\subsection{Sequence of Games}

The sequence of games represents the overall strategy for completing the proof.  In the case of PRF Encryption, we want to show that the probability that the adversary will correctly guess the randomly chosen ``world" is close to \nicefrac{1}{2}.  We accomplish this by instantiating the IND-CPA security definition with the construction, and then transforming this game, little by little, until we have a game in which this probability is exactly \nicefrac{1}{2}.  Each transformation may add some concrete value to the bounds, and we want to ensure that the sum of these values is small.

\begin{figure}[h]

\centering

\begin{tikzpicture}[node distance=0.75cm,
  game/.style={rectangle,rounded corners,draw,minimum height=0.6cm, font=\sffamily\normalsize\bfseries},
  equiv/.style={font=\sffamily\large\bfseries},
  outer/.style={rectangle,draw,font=\sffamily\Large\bfseries}]

  \node[outer] (R1){
    \begin{tikzpicture}
      \node[game] (IND_CPA_G) {IND\_GPA\_G};
      \node[equiv] (IND_CPA_G1) [right of=IND_CPA_G, xshift=0.7cm] {=};
      \node[game] (G1) [right of=IND_CPA_G1] {G1};
    \end{tikzpicture}
  };

  \node[equiv] (G1_G2) [below of=R1, xshift=1.0cm] {$\approx_\text{PRF\_Advantage}$};
  \node[game] (G2) [below of=G1_G2, xshift=-1.0cm, minimum width=2cm] {G2};

  \node[equiv] (G2_G3) [below of=G2, xshift=1.4cm] {$\approx_\text{Random List Collision}$};
  \node[game] (G3) [below of=G2_G3, xshift=-1.4cm, minimum width=2cm] {G3};

  \node[equiv] (G3_G4) [below of=G3, xshift=0.95cm] {$=_\text{One Time Pad}$};

  \node[outer] (R2) [below of=G3_G4, xshift=-0.95cm]{
    \begin{tikzpicture}    
      \node[game] (G4) {G4};
      \node[equiv] (G4_G5) [right of=G4] {=};
      \node[game] (G5) [right of=G4_G5] {G5};
      \node[equiv] (G5_Half) [right of=G5] {=};
      \node[game] (Half) [right of=G5_Half] {\nicefrac{1}{2}};
    \end{tikzpicture}
  };

\end{tikzpicture}

\caption{Sequence of Games Diagram}
\label{games_diagram}
\end{figure}
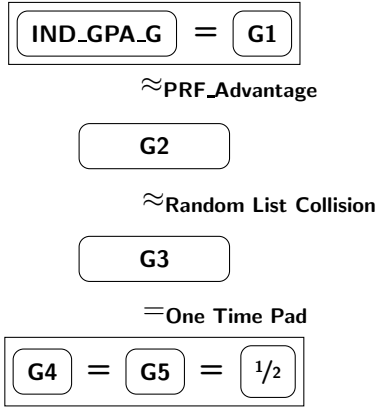

The diagram in Figure \ref{games_diagram} shows the entire sequence of games, as well as the relationship between each pair of games in the sequence.  In this diagram, two games are related by $=$ if they are identical, and by $\approx$ if they are close.  When the equivalence is non-trivial, the diagram gives an argument for the equivalence, which implies a bound on the distance between the games when they are not equal.  A detailed description of each game transformation follows:

\begin{enumerate}[topsep=0pt,partopsep=8pt,parsep=1pt,itemsep=1pt]
\item Instantiate the IND-CPA definition with the construction.  Unfold definitions and simplify.  (Listing \ref{Game1}) 
\item Apply the PRF definition to replace the PRF with a random function.  (Listing \ref{Game2}) 
\item Replace the random function output used to encrypt the challenge ciphertext with a bit vector selected completely at random.  With overwhelming probability, the adversary does not notice this change.  (Listing \ref{Game3})
\item We have modified the game to the point that encryption of the challenge plaintext is by one-time pad.  So we can replace the ciphertext with a randomly-chosen value.  (Listing \ref{Game4}) 
\item Now the ciphertext is independent from the plaintext, and thus independent from the random bit that was used to select the ``world."  This means we can move this coin flip to \textit{after} the adversary guesses which world he is in.  In this game, it is obvious that the probability that the adversary guesses the correct outcome of the coin flip is exactly one half.  (Listing \ref{Game5})   
\end{enumerate}
\vspace*{-\baselineskip}

\begin{lstlisting}[float=h, captionpos=b, label=Game1, caption=Game 1]

Definition G1 :=
  key <-$ PRFE_KeyGen;
  [b, _] <-$2
  (
    [p0, p1, s_A] <--$3 A1;
    b <--$$ {0, 1};
    pb <- if b then p1 else p0;
    c <--$$ PRFE_Encrypt key pb;
    b' <--$ (A2 s_A c);
    $ ret (eqb b b')
  )
  (PRFE_EncryptOracle key) tt;
  ret b.

\end{lstlisting}

\begin{lstlisting}[float=h, captionpos=b, label=Game2, caption=Game 2]
Definition PRFE_RandomFunc := @randomFunc 
  (Bvector eta) (Bvector eta) ({0,1}^eta) _.

Definition RF_Encrypt s p :=
  r <-$ {0, 1} ^ eta;
  [pad, s] <-$2 PRFE_RandomFunc s r;
  ret (r, p xor pad, s).
    
Definition G2 :=
  [a, o] <-$2 A1 (RF_Encrypt) nil;
  [p0, p1, s_A] <-3 a;
  b <-$ {0, 1};
  pb <- if b then p1 else p0;
  [c, o] <-$2 RF_Encrypt o pb;
  [b', o] <-$2 (A2 s_A c) RF_Encrypt o;
  ret (eqb b b').
\end{lstlisting}

\begin{lstlisting}[float=h, captionpos=b, label=Game3, caption=Game 3]

Definition G3 :=
  [a, o] <-$2 A1 (RF_Encrypt) nil;
  [p0, p1, s_A] <-3 a;
  b <-$ {0, 1};
  pb <- if b then p1 else p0;
  r <-$ {0, 1}^eta;
  pad <-$ {0, 1}^eta;
  c <- (r, pb xor pad);
  [b', o] <-$2 (A2 s_A c) RF_Encrypt o;
  ret (eqb b b').
\end{lstlisting}

\begin{lstlisting}[float=h, captionpos=b, label=Game4, caption=Game 4]
Definition G4 :=
  [a, o] <-$2 A1 (RF_Encrypt) nil;
  [p0, p1, s_A] <-3 a;
  b <-$ {0, 1};
  pb <- if b then p1 else p0;
  r <-$ {0, 1}^eta;
  pad <-$ {0, 1}^eta;
  c <- (r, pad);
  [b', o] <-$2 (A2 s_A c) RF_Encrypt o;
  ret (eqb b b').
\end{lstlisting}

\begin{lstlisting}[float=h, captionpos=b, label=Game5, caption=Game 5]
Definition G5 :=
    [a, o] <-$2 A1 (RF_Encrypt) nil;
    [p0, p1, s_A] <-3 a;
    r <-$ {0, 1}^eta;
    pad <-$ {0, 1}^eta;
    c <- (r, pad);
    [b', o] <-$2 (A2 s_A c) RF_Encrypt o;
    b <-$ {0, 1};
    ret (eqb b b').
\end{lstlisting}

\subsection{Equivalence Proofs}
\label{equivalence_proofs_section}

The next step is to prove the appropriate sort of equivalence between each pair of games in the sequence.  In the case of PRF Encryption, the goal is to show that the distance between the IND-CPA game and \nicefrac{1}{2} is very small, and we accomplish this by showing that each pair of games in the sequence is either identical or ``close." 

The first step is to show that Game 1 (Listing \ref{Game1}) really is the IND-CPA game instantiated with this encryption scheme.  This fact (Listing \ref{Equiv1}) is obvious, and the proof can be completed using Coq's \texttt{reflexivity} tactic.  \texttt{A1} and \texttt{A2} are parameters representing the procedures of the adversary against the encryption scheme.  In the statement of this theorem, \texttt{==} is equality for rational numbers.  This equality is registered with Coq's setoid system to enable tactics such as \texttt{reflexivity} and rewriting.           

\begin{lstlisting}[float=h, captionpos=b, label=Equiv1, caption=Equivalence of the Security Definition and Game 1]
Theorem G1_equiv : 
  Pr[IND_CPA_SecretKey_G PRFE_KeyGen PRFE_Encrypt A1 A2] == Pr[G1].
\end{lstlisting}

Next we show that the distance between Games 1 and 2 is exactly the advantage of some adversary against a PRF.  The adversary against the PRF (Listing \ref{PRF_A}) is constructed from \texttt{A1} and \texttt{A2}.  \texttt{PRFE\_Encrypt\_OC} is an encryption oracle that interacts with the PRF as an oracle.  \texttt{PRF\_A} provides this encryption oracle to \texttt{A1} and \texttt{A2} using the \texttt{OC\_Run} operation.    

\begin{lstlisting}[float=h, captionpos=b, label=PRF_A, caption=The Constructed Adversary Against the PRF]

Definition PRFE_Encrypt_OC (x : unit)
  (p : Plaintext) : OracleComp (Bvector eta) 
  (Bvector eta) (Ciphertext * unit) :=
  r <--$$ {0, 1} ^ eta;
  pad <--$ OC_Query r;
  $ (ret (r, p xor pad, tt)).

Definition PRF_A : OracleComp (Bvector eta) 
  (Bvector eta) bool :=
  [a, n] <--$2 OC_Run A1 PRFE_Encrypt_OC tt;
  [p0, p1, s_A] <-3 a;
  b <--$$ {0, 1};
  pb <- if b then p1 else p0;
  r <--$$ {0, 1} ^ eta;
  pad <--$ OC_Query r;
  c <- (r, pb xor pad);
  z <--$ OC_Run (A2 s_A c) PRFE_Encrypt_OC n;
  [b', _] <-2 z;
  $ ret (eqb b b').

\end{lstlisting}

To prove the ``closeness" of Games 1 and 2, first we prove that the interaction between \texttt{PRF\_A} and the PRF oracle is equivalent to Game 1 (Theorem \ref{Equiv2_G1}).
Then we prove that the interaction between \texttt{PRF\_A} and the random function oracle is equivalent to Game 2 (Theorem \ref{Equiv2_G2}).  Finally we apply the results of parts 1 and 2 and unify with the definition of a PRF (Theorem \ref{G1_G2_Close}).

\begin{lstlisting}[float=h, captionpos=b, label=Equiv2_G1, caption=Equivalence PRF\_A and G1]
Theorem G1_PRF_A_equiv : 
  Pr[k <-$ {0, 1}^ eta; 
    [b, _] <-$2 PRF_A (f_oracle k) tt; 
    ret b] == Pr[G1].
\end{lstlisting}

\begin{lstlisting}[float=h, captionpos=b, label=Equiv2_G2, caption=Equivalence PRF\_A and G2]
Theorem G2_PRF_A_equiv : 
  Pr[[b, _] <-$2 PRF_A PRFE_RandomFunc nil; 
    ret b] == Pr[G2].
\end{lstlisting}

\begin{lstlisting}[float=h, captionpos=b, label=G1_G2_Close, caption=Closeness of Game 1 and Game 2]
Theorem G1_G2_close : | Pr[G1] - Pr[G2] | == 
  PRF_Advantage ({0, 1}^eta) ({0, 1}^eta) 
    f PRF_A.
\end{lstlisting}

To prove Theorems \ref{Equiv2_G1} and \ref{Equiv2_G2}, we mostly perform simple manipulations such as applying the denotational semantics of \texttt{OracleComp}, inlining, and removing identical statements at the beginning of the game.  In both of these proofs, the adversary is interacting with two different, but observationally equivalent, oracles.  So we use the program logic and Theorem \ref{OracleEquiv} to prove that these interactions produce equivalent results.  

Next we show that Games 2 and 3 are ``close" by demonstrating that these games are ``identical until bad" in the sense of Theorem \ref{IdenticalUntilBad}.  The ``bad" event of interest is the event that the randomly-generated PRF input used to encrypt the challenge plaintext (\texttt{r} in Game 2) is also used to encrypt some other value during the interaction between the adversary and the encryption oracle.  There are two separate adversary procedures, and each one is capable of encountering \texttt{r} during its interaction with the oracle.  So we divide this proof into two parts, one for each adversary procedure, where each part includes an ``identical until bad" argument.  In the first step, we produce the \texttt{pad} value randomly (without using the random function), but then add an entry for \texttt{r} and \texttt{pad} to the state of the random function.  In the second step, we produce \texttt{pad} randomly, and do not add an entry to the random function. 

To get an expression for the probability of the ``bad"event, we assume natural numbers $q_1$ and $q_2$, and that \texttt{A1} performs at most $q_1$ queries and \texttt{A2} performs at most $q_2$ queries.  FCF includes a library module called \texttt{RndInList} that includes general-purpose arguments related to the probability of encountering a randomly selected value in a list of a certain length, and the probability of encountering a certain value in a list of randomly-generated elements of a certain length.  By combining the ``identical until bad" proofs with these arguments to get expressions bounding the probabilities of the bad events, we obtain the result of Listing \ref{G2_G3_Close}.

\begin{lstlisting}[float=h, captionpos=b, label=G2_G3_Close, caption=Closenes of Game 2 and Game 3]
Theorem G2_G3_close : 
  | Pr[G2] - Pr[G3] | <= 
    q1 / (2 ^ eta) + q2 / (2 ^ eta).
\end{lstlisting}    

The next step is to use a one-time-pad argument to replace the challenge ciphertext with a randomly-chosen value.  The library contains a generic one-time-pad argument that we can apply here.  We transform this game into an equivalent game that unifies with the one-time-pad argument, then we apply the argument to get the result shown in Listing \ref{Equiv4}.      

\begin{lstlisting}[float=h, captionpos=b, label=Equiv4, caption=Equivalence of Game 3 and Game 4]
Theorem G3_G4_equiv : Pr[G3] == Pr[G4].
\end{lstlisting}

Now that the ciphertext is independent of the challenge bit, we produce a new game by moving the sampling of the challenge bit to the end of the game.  To prove this fact (Listing \ref{Equiv5}), we simply unfold the required definitions, skip over all of the identical pairs of statements at the beginning of the proof, then swap the order of independent statements in the game on the left in order to make these statements align with the identical statements in the game on the right.  

\begin{lstlisting}[float=h, captionpos=b, label=Equiv5, caption=Equivalence of Game 4 and Game 5]
Theorem G4_G5_equiv : 
  Pr[G4] == Pr[G5].
    unfold G4, G5.
    do 3 (comp_skip; comp_simp; comp_swap_l).
    comp_skip; comp_simp.
    reflexivity.
  Qed.
\end{lstlisting}

Finally, we develop the proof that the adversary wins Game 5 with probability exactly $\nicefrac{1}{2}$.  This proof (Listing \ref{Half5}) proceeds by discarding all of the initial statements in the game using the \texttt{comp\_irr\_l} tactic.  Note that this tactic produces an obligation to prove that the statement being discarded is a well-formed computation, which can be discharged with the tactic \texttt{wftac}.  Then what remains is a very simple game, and \texttt{dist\_compute} can automatically compute the probability that this game returns \textit{true}.  

\begin{lstlisting}[float=h, captionpos=b, label=Half5, caption=Probability of Winning Game 5]
Theorem G5_one_half : 
  Pr[G5] == 1/2.
    do 4 comp_irr_l; wftac.  
    dist_compute.  
Qed.
\end{lstlisting}

By combining the equivalences of each pair of intermediate games, we get the final concrete security result shown in Listing \ref{ConcreteResult}.  It is important to note that the statement of this theorem does not reference any of the intermediate games.  The sequence of games was only a tool that we used to get the final result, and this sequence does not need to be inspected in order to trust the result.  

\begin{lstlisting}[float=h, captionpos=b, label=ConcreteResult, caption=Concrete Security Result]
Theorem PRFE_IND_CPA_concrete : 
  IND_CPA_SecretKey_Advantage PRFE_KeyGen PRFE_Encrypt A1 A2 <=
  PRF_Advantage ({0, 1}^eta) ({0, 1}^eta) 
    f PRF_A + (q1 / 2^eta + q2 / 2^eta).
\end{lstlisting}

The concrete security result in Listing \ref{ConcreteResult} may be sufficient for many purposes.  We have an expression describing the advantage of the adversary, and we can inspect this expression to see whether this advantage is sufficiently small.  We also must inspect the definition of the adversary \texttt{PRF\_A}, which appears in this result, and ensure that this adversary is ``efficient" according to the desired complexity class.  Next we will show how to derive an asymptotic security result based on this concrete result.  A benefit of proving asymptotic security is that this proof removes the requirement to inspect the constructed adversary and the expression describing the adversary's advantage.   

\subsection{Asymptotic Security Definitions}

Now we give the asymptotic security definitions for PRFs and IND-CPA encryption.  These definitions are parameterized by an admissibility predicate as described in Section \ref{EfficiencySection}.  The IND-CPA definition accepts two admissibility predicates -- one for each adversary procedure.  

The asymptotic security definition for a PRF is given in Listing \ref{PRF_Def}.  In this definition, \texttt{RndKey}, \texttt{RndR}, and \texttt{f} are \texttt{nat}-indexed families of procedures.  Similarly in the IND-CPA definition (Listing \ref{IND_CPA_Def}), \texttt{KeyGen} and \texttt{Encrypt} are \texttt{nat}-indexed families of procedures.  Both of these definitions are claims over all admissible \texttt{nat}-indexed adversary families.  Note that both definitions reuse the expressions provided in the concrete security definitions.  This style provides a convenient method for developing an asymptotic security proof from a concrete security proof.   

\begin{lstlisting}[float=h, captionpos=b, label=PRF_Def, caption= Definition of a PRF]
Variable D R Key : nat -> Set.
Variable RndKey : forall n, Comp (Key n).
Variable RndR : forall n, Comp (R n).
Variable f : forall n, Key n -> D n-> R n.

Definition PRF :=
  forall (A : \forall n, OracleComp (D n) (R n) 
    bool), admissible_A A -> 
    negligible (fun n => PRF_Advantage 
      (RndKey n) (RndR n) (@f n) (A n)).

\end{lstlisting}

\begin{lstlisting}[float=h, captionpos=b, label=IND_CPA_Def, caption= Definition of IND-CPA Encryption]
Variable Plaintext Ciphertext Key State : 
  nat -> Set.
Variable KeyGen : forall n, Comp (Key n).
Variable Encrypt : forall n, Key n ->  
  Ciphertext n -> Comp (Plaintext n).

Definition IND_CPA_SecretKey :=
  forall (State : nat -> Set)
  (A1 : forall n, OracleComp (Plaintext n) 
    (Ciphertext n) 
    (Plaintext n * Plaintext n * State n))
  (A2 : forall n, State n -> Ciphertext n -> OracleComp (Plaintext n) (Ciphertext n) bool),
    admissible_A1 A1 ->
    admissible_A2 A2 ->
    negligible 
      (fun n => IND_CPA_SecretKey_Advantage 
      (KeyGen n) (@Encrypt n) (A1 n) (A2 n) ).

\end{lstlisting}

\subsection{Efficiency of Constructed Adversaries}

The first step in proving an asymptotic security result is to view each constructed adversary in the concrete proof as a \texttt{nat}-indexed family of adversaries, and prove that this family is ``efficient" as defined by some complexity class.  In the PRF Encryption proof, we use the non-uniform polynomial time complexity class described in Section \ref{EfficiencySection}.  Because this class includes a concrete cost model, we begin with a proof of the concrete cost of each constructed adversary procedure.  

We begin by assuming costs for \texttt{A1} and \texttt{A2}.  \texttt{A1\_cost} is a function describing the cost of \texttt{A\_1}.  \texttt{A2\_cost\_1} is a number describing how much it costs for \texttt{A2} to compute an \texttt{OracleComp} that is closed over a state and a ciphertext.  Then \texttt{A2\_cost\_2} is a function describing the cost of executing this \texttt{OracleComp}.  Given these assumptions, we can give a cost to \texttt{PRF\_A} as shown in Listing \ref{PRF_A_Cost}.  In the statement of this theorem, \texttt{oc\_cost}, \texttt{comp\_cost}, and \texttt{cost} are the cost models for \texttt{OracleComp}, \texttt{Comp}, and Coq functions, respectively.  Note that this cost model is overly conservative and some costs are counted multiple times.    

\begin{lstlisting}[float=h, captionpos=b, label=PRF_A_Cost, caption=Cost of Constructed Procedure PRF\_A]	
Theorem PRF_A_cost : 
  oc_cost cost (comp_cost cost) PRF_A
    (fun x => (A1_cost (x + (5 * eta))) + 
    (A2_cost_2 (x + (5 * eta))) + 
     x + 5 * A2_cost_1 + 6 + 7 * eta).
\end{lstlisting}

This proof is completed by repeatedly applying the rule of the cost model that is relevant to the term in the goal, which is a highly syntax-directed operation that can be mostly automated.  Once all these syntax-directed rules are applied, the developer is obligated to prove that the expression obtained in this process is equal to (or less than) the expression in the statement of the theorem.  In this last step of the proof, automated tactics such as \texttt{omega} are very useful.    

\subsection{Asymptotic Security Proof}

The final step in the proof is to show that the security definition shown in Listing \ref{IND_CPA_Def} holds on this construction as long as \texttt{f} is a PRF as defined in Listing \ref{PRF_Def}.  The statement of this fact is shown in Listing \ref{PRFE_Security}.  Note that \texttt{admissible\_oc} and \texttt{admissible\_oc\_func\_2} are the admissibility predicates for \texttt{OracleComp} and for functions with two arguments that produce an \texttt{OracleComp} defined in the simple complexity class described in Section \ref{EfficiencySection}.  

\begin{lstlisting}[float=h, captionpos=b, label=PRFE_Security, caption=Asymptotic Security of PRF Encryption]	
Theorem PRFE_IND_CPA : 
  PRF Rnd Rnd f (admissible_oc cost) -> 
  IND_CPA_SecretKey
    PRFE_KeyGen (fun n => PRFE_Encrypt (@f n))
    (admissible_oc cost) 
    (admissible_oc_func_2 cost).
\end{lstlisting}

The primary obligation of this proof is to show that the function defining the advantage of any admissible family of adversaries against this encryption scheme is a negligible function.  The fact that this adversary family is admissible allows us to use the result of Listing \ref{PRF_A_Cost}, along with other facts, to conclude that the constructed adversary family against the PRF is admissible.  In the course of this proof, we must show that the expression implied by Figure \ref{PRF_A_Cost} is at most polynomial in $\eta$ if x is at most polynomial in $\eta$ and all the costs related to \texttt{PRF\_A1} and \texttt{PRF\_A2} are at most polynomial in $\eta$.  This fact is proven using the provided theory of polynomial functions (Section \ref{AsymptoticSection}).
  
From the admissibility of the constructed adversary, and from the fact the \texttt{f} is a PRF against all admissible adversaries, we can conclude that the constructed adversary's advantage against the PRF is negligible.  The advantage of this adversary against the PRF is one of the terms that appears in the bounds of the concrete result (Listing \ref{ConcreteResult}).  The other term is $q_1 / 2 ^ \eta + q_2 / 2 ^ \eta$, where $q_1$ and $q_2$ are the number of oracle queries performed by the two adversary procedures.  The admissibility predicates ensure that each adversary only performs a polynomial number of queries, so $q_1$ and $q_2$ must be polynomial in $\eta$, and this expression is negligible in $\eta$.  So the advantage of the adversary against this encryption scheme is the sum of two negligible functions, and is therefore negligible.  

The entire proof of security for this encryption scheme requries approximately 1500 lines of Coq code, of which about 700 lines are specification (including 100 lines of cryptographic definitions and intermediate games) and 800 lines are proof.  The proof incorporates another 500 lines of code for the reusable arguments (\emph{e.g.}, the one-time pad argument).  We expect that a skilled Coq developer could complete such a proof in a matter of days (though he may require the help of a cryptographer to develop the sequence of games and high-level arguments).  

\section{Evaluation}
\label{EvaluationSection}

This section attempts to evaluate FCF against the design goals listed in Section \ref{properties_section}, and to contrast with both CertiCrypt and EasyCrypt.  

All three of these frameworks provide concrete bounds, so this criterion is not discussed further.  
And, all three frameworks use a relatively familiar syntax for security definitions and constructions.  We believe that, based on our experience working with cryptographers, they can easily understand these definitions (\emph{e.g.}, Listing \ref{PRF_Def_Concrete}) after spending a few minutes familiarizing themselves with the notation.   

Regarding proof automation, FCF lies somewhere between CertiCrypt and EasyCrypt.  EasyCrypt achieves a significant level of automation by using SMT solvers to discharge simple logical goals, but higher-level goals still need to be addressed manually by applying tactics.  FCF achieves a similarly high level of automation through the use of existing and custom Coq tactics.  These tactics are not as powerful as modern SMT solvers, so the developer may need to manually address some goals in FCF that would be discharged automatically in EasyCrypt.  However, the semantics of programs in FCF is computational, so Coq is able to immediately compute an expression describing the probability distribution for any program.  This allows some simple equivalences to be discharged immediately using this computation and FCF's \texttt{dist\_compute} tactic.  

Regarding trust in \emph{extensional properties}, FCF and CertiCrypt are foundational, meaning that the program logic is constructed definitionally from the semantics.  In contrast, EasyCrypt relies upon a set of axioms for its program logic.  EasyCrypt also relies on the correctness of the EasyCrypt front end and the Why3 verification generator, whereas FCF and CertiCrypt only depend on the Coq type checker.  EasyCrypt provides no support for reasoning about \emph{intensional properties} like execution time, whereas CertiCrypt and FCF do, though FCF provides this suport using a trusted set of axioms.    

EasyCrypt and CertiCrypt are based on simply-typed, first-order languages.  This design makes it difficult to directly support abstraction, extension, and reuse, though these frameworks include elements which support these goals to some extent.  In contrast, FCF uses a shallow embedding and the advanced features of Coq, such as dependent types, modules, notation, and higher-order functions, to support abstraction, extensiblity, and reuse.  We believe that having such a rich language for describing games and assumptions is critical for scaling to larger protocols.  

FCF supports code generation with a semantics that is proven to be equivalent to the semantics used to reason about the probabilistic behavior of programs.  That is, a program extracted from an FCF model is guaranteed to produce the correct probability distribution when the input bits provided to it are uniformly distributed, assuming the extraction mechanism of Coq preserves meaning.  There has been some initial work in producing implementations that correspond to EasyCrypt models, but there is no formal relationship between the semantics of the implementation and the semantics used to reason about the model.

\section{Related Work}
\label{RelatedWorkSection}
There has been a large amount of work in the area of verifying cryptographic schemes in recent years.  In this section we will describe some of this related work, focusing on systems that attempt to establish security in the computational model.  CertiCrypt~\cite{Zanella:2009:POPL} and EasyCrypt~\cite{Zanella:2011:CRYPTO} have been thoroughly discussed previously in this paper.  


There are several other examples of frameworks for cryptographic security proofs implemented within proof assistants.  The most similar work is that of Nowak~\cite{cryptoeprint:2007:199}, who was the first to develop proofs of cryptography in Coq using a shallow embedding in which programs have probability distributions as their denotations.  FCF builds on this work by adding more tools for modeling and reasoning such as procedures with oracle access (Section \ref{LangSection}), a program logic (Section \ref{ProgramLogic}), and asymptotic reasoning (Section \ref{AsymptoticSection}).  

The work of \cite{Affeldt:2007:FPP:1779394.1779408} is a Coq library utilizing a deeply-embedded imperative programming language.  This library is a predecessor to CertiCrypt, and it includes some important elements that were later adopted by CertiCrypt.  Notably, the probabilistic programming language in this work is given a semantics in which program states are distributions, and the semantics describes how these distributions are transformed by each command in the language.  CertiCrypt and EasyCrypt extended this work by adding language constructs such as oracles and unrestricted loops, and well as reasoning tools such as the Probabilistic Relational Hoare Logic.  

Verypto~\cite{berg13thesis} is a fully-featured framework built on Isabelle~\cite{Nipkow-Paulson-Wenzel:2002} that includes a deep embedding of a functional programming language.  To allow state information to remain hidden from adversaries, Verypto provides ML-style references, in contrast to the oracle system provided by FCF.  To date, Verypto has only been used to prove the security of simple constructions, but this work uses an interesting approach that deserves more exploration.   

CryptoVerif~\cite{BlanchetCSF07} is a tool based on a concurrent, probabilistic process calculus that is only able to prove properties related to secrecy and authenticity.  CryptoVerif is highly automated to the extent that it will even attempt to locate intermediate games, and so proof development in CryptoVerif requires far less effort compared to FCF or EasyCrypt.  However, there are a large number of proofs that could be completed in FCF or EasyCrypt that are impossible in CryptoVerif due to its specialized nature.  

Refinement types~\cite{Bengtson08refinementtypes} have been used by Fournet et al~\cite{DBLP:conf/ccs/FournetKS11} to develop proofs of security for cryptographic schemes in the computational model.  In this system, a security property is specified as an ideal functionality (in the sense of the real/ideal paradigm), and proofs are completed using the ``sequence of games" style in the asymptotic setting.  This approach allows the proofs of security to be fairly simple, but no concrete security claims are proved, so it may be difficult to make practical claims based on such a proof.  

Computational soundness~\cite{Abadi:2000:RTV:647318.723498} provides another mechanism for verifying cryptographic schemes.  This approach attempts to derive security in the computational model from security in the symbolic model by showing that any likely execution trace in the computational model also exists in the symbolic model.  It is possible to mechanize such a proof as described in \cite{backes08computational}.  This approach is limited to classes of schemes for which computational soundness results have been discovered.  Another limitation with this approach is that it can only produce proofs in the asymptotic setting---there is no way to prove concrete security claims.  

Protocol Composition Logic (PCL)~\cite{datta2007protocol} provides a logic and proof system for verifying cryptographic schemes in the symbolic model.  The system is based on a process calculus and allows reasoning about the results of individual protocol steps.  More recent work~\cite{Datta:2005:PPS:2104063.2104066} has extended this logic to allow for proofs in the computational model.  In computational PCL, formulas are interpreted against probability distributions on traces and a formula is true if it holds with overwhelming probability.  This approach is similar to computational soundness in that low-probability traces are ignored, and proofs of concrete security claims are impossible.  

\section{Conclusion and Future Work}

Our contribution is a complete mechanized framework for specifying and checking cryptographic proofs within a proof assistant.  Our framework compares favorably to the current state of the art, and provides many new benefits, such as extensibility through a foundational approach, a powerful language for describing schemes, and the ability to extract excutable code.  Of course, whether these benefits apply at scale is still an open question, and thus a key direction for us is to prove security for an even wider range of standard constructions, as well as novel cryptographic schemes.  
In particular, we have proven the security of the ``tuple set" construction of \cite{cash13sse}, and we intend to continue developing this work into a proof of a searchable symmetric encryption scheme.

The biggest limitation of FCF is that it currently lacks definitions for many cost models and complexity classes that are commonly used in cryptography.  We hope to develop more cost models and complexity classes, including a complexity class describing (uniform) probabilistic polynomial time Turing machines.


\bibliographystyle{abbrvnat}
\bibliography{fcf}

\end{document}